\documentclass[a4paper,12pt]{article}
\usepackage{graphicx,amsmath,amssymb}
\usepackage{fullpage,mathptmx}
\newtheorem{theorem}{Theorem}
\newtheorem{lemma}{Lemma}
\DeclareMathOperator{\KS}{\textit{C}}
\DeclareMathOperator{\poly}{\textrm{poly}}
\begin{document}
\title{Kolmogorov complexity and cryptography}
\author{Andrej A.~Muchnik\thanks{This paper contains some results of An.A.~Muchnik (1958--2007) reported in his talks at the Kolmogorov seminar (Moscow State Lomonosov University, Math. Department, Logic and Algorithms theory division, March 11, 2003 and April 8, 2003) but not published at that time. These results were stated (without proofs) in the joint talk of Andrej Muchnik and Alexei Semenov at Dagstuhl Seminar 03181, 27.04.2003--03.05.2003. This text was prepared by Alexey Chernov and Alexander Shen in 2008--2009.}}
\date{}
\maketitle
\begin{abstract}
  We consider (in the framework of algorithmic information theory) questions of the following type: construct a message that contains different amounts of information for recipients that have (or do not have) certain a priori information.
  
Assume, for example, that the recipient knows some string $a$, and we want to send her some information that allows her to reconstruct some string $b$ (using $a$). On the other hand, this information alone should not allow the eavesdropper (who does not know $a$) to reconstruct $b$. It is indeed possible (if the strings $a$ and $b$ are not too simple).

Then we consider more complicated versions of this question. What if the eavesdropper knows some string $c$? How long should be our message? We provide some conditions that guarantee the existence of a polynomial-size message; we show then that without these conditions this is not always possible.

\end{abstract}

\section{Non-informative conditional descriptions}

In this section we construct (for given strings $a$ and $b$ that satisfy some conditions) a string $f$ that contains enough information to obtain $b$ from $a$, but does not contain any information about $b$ in itself (without $a$), and discuss some generalizations of this problem.

\subsection*{Uniform and non-uniform complexity}

Let us start with some general remarks about conditional descriptions and their complexity. Let $X$ be a set of binary strings, and let $y$ be a string. Then $\KS(X\to y)$ can be defined as the minimal length of a program that maps every element of $X$ to $y$. (As usually, we fix some optimal programming language. We can also replace minimal length by minimal complexity.) Evidently,
         $$
       \KS(X\to y) \ge \max_{x\in X}\KS(y|x)
         $$
(if a program $p$ works for all $x\in X$, it works for every $x$), but the reverse inequality is not always true. It may happen that the ``uniform'' complexity of the problem $X\to y$ (left hand side) is significantly greater than the ``nonuniform'' complexity of the same problem (right hand side).

To prove this, let us consider an incompressible string $y$ of length $n$ and let $X$ be the set of all strings $x$ such that $\KS(y|x)<n/2$. Then the right hand side is bounded by $n/2$ by construction. Let us show that left hand side is greater than $n-O(\log n)$.  Indeed, let $p$ be a program that outputs $y$ for every input $x$ such that $\KS(y|x)<n/2$. Among those $x$ there are strings of complexity $n/2+O(\log n)$ and together with $p$ they are enough to obtain $y$, therefore $\KS(y|p) \le n/2+O(\log n)$. Therefore, there exists a string $e$ of length $O(\log n)$ such that $\KS(y|\langle p,e\rangle)<n/2$. Then, by our assumption, $p(\langle p, e\rangle)=y$ and therefore the complexity of $p$ is at least $n-O(\log n)$.

\textbf{Remark}. In this example the set $X$ can be made finite if we restrict ourselves to strings of bounded length, say,  of length at most $2n$.

\subsection*{Complexity of the problem $(a\to b)\to b$}

The example above shows that uniform and nonuniform complexities could differ significantly. In the next example they coincide, but some work is needed to show that they coincide.

Let $a$ and $b$ be binary strings. By $(a\to b)$ we denote the set of all programs that transform input $a$ into output $b$. It is known~\cite{shen-ver} that
	$$
\KS((a\to b)\to b) = \min (\KS(a),\KS(b))+O(\log N)
	$$
for any two strings $a,b$ of length at most $N$. It turns out that a stronger version of this statement (when the uniform complexity is replaced by a non-uniform one) is also true:

\begin{theorem}
	\label{th:1}
For every two strings $a$ and $b$ of length at most $N$ there exists a program $f$ that maps $a$ to $b$ such that
	$$
\KS(b|f)=\min\{\KS(a),\KS(b)\} + O(\log N)
    $$

\end{theorem}

\textbf{Proof}.
Note that $\le$-inequality is obviously true for any program $f$ that maps $a$ to $b$. Indeed, having such a function and any of the strings $a$ and $b$, we can reconstruct $b$.

Let us prove that the reverse inequality is true for some function $f$ that maps $a$ to $b$. We restrict ourselves to total functions defined on the set of all strings of length at most $n$ and whose values also belong to this set, so such a function is a finite object and conditional complexity with respect to $f$ is defined in a natural way. Note also that  (up to $O(\log N)$ precision) it does not matter whether we consider $f$ as an explicitly given finite object or as a program, since (for known $N$) both representations can be transformed to each other.

Let $m$ be the maximum value of $\KS(b|f)$ for all functions (of the type described) that map $a$ to $b$. We need to show that one of the strings $a$ and $b$ has complexity at most $m+O(\log N)$. This can be done as follows.

Consider the set $S$ of all pairs $\langle a', b'\rangle$ where $a'$ and $b'$ are strings of length at most $N$ that have the following property: \emph{$\KS(b'|f)\le m$ for every total function $f$ whose arguments and values are string of length at most $N$ and $f(a')=b'$}. By the definition of $m$, the pair $\langle a,b\rangle$ belongs to $S$.

The set $S$ can be effectively enumerated given $m$ and $N$. Let us perform this enumeration and delete pairs whose first  or second coordinate was already encountered (as the first/second coordinate of some other undeleted pair during the enumeration); only ``original'' pairs with two ``fresh'' components are placed in $\tilde S$.  This guarantees that $\tilde S$ is a graph of a bijection. The pair $\langle a, b\rangle$ is not necessarily in $\tilde S$; however, some other pair with the first component $a$ or with the second component $b$ is in $\tilde S$ (otherwise nothing prevents $\langle a,b\rangle$ from appearing in $\tilde S$).

Since $\tilde S$ can also be effectively enumerated (given $m$ and $N$),  it is enough to show that it contains $O(2^m)$ elements (then the ordinal number of the above-mentioned pair describes either $a$ or $b$). 

To show this, let us extend $\tilde S$ to the graph of some bijection~$g$. If some $\langle a',b'\rangle\in \tilde S$, then $g(a')=b'$ and therefore $\KS(b'|g)\le m$ by construction (recall that $\tilde S$ is a subset of $S$). Therefore, $\tilde S$ contains at most $O(2^m)$ different values of $b'$, but $\tilde S$ is a bijection graph. (End of proof.)

\subsection*{Cryptographic interpretation}

Theorem~\ref{th:1} has the following ``cryptographic'' interpretation. We want to transmit some information (string $b$) to an agent that already knows some ``background'' string $a$ by sending some message $f$. Together with $a$ this message should allow the agent to reconstruct $b$. At the same time we want $f$ to carry minimal information about $b$ for a ``non-initiated'' listener, i.e., the complexity $\KS(b|f)$ should be maximal. This complexity cannot exceed $\KS(b)$ for evident reasons and cannot exceed $\KS(a)$ since $a$ and $f$ together determine $b$. Theorem~\ref{th:1} shows that this upper bound can be reached for an appropriate $f$.

Let us consider a relativized version of this result that also has a natural cryptographic interpretation. Assume that non-initiated listener knows some string $c$. Our construction (properly relativized) proves the existence of a function $f$ that maps $a$ to $b$ such that 
	$$
\KS(b|f,c) \approx \min (\KS(a|c), \KS(b|c)).
	$$
This function has minimal possible amount of information about $b$ for people who know $c$. More formally, the following statement is true (and its proof is a straightforward relativization of the previous argument):

\begin{theorem}
	\label{th1:relativized}
Let $a,b,c$ be strings of length at most $N$. Then there exists a string $f$ such that:
	
	\textup{(1)}~$\KS(b|a,f)=O(\log N)$\textup;
	
	\textup{(2)}~$\KS(b|c,f)=\min\{\KS(a|c),\KS(b|c)\} + O(\log N)$.

\end{theorem}

The claim~(1) says that for recipients who know $a$ the message $f$ is enough to reconstruct $b$; the claim~(2) says that for the recipients who know only $c$ the message $f$ contains minimal possible information about~$b$.

\textbf{Remark}. One may try to prove Theorem~\ref{th:1} as follows: let $f$ be the shortest description of $b$ when $a$ is known; we may hope that it does not contain ``redundant'' information. However, this approach does not work: if $a$ and $b$ are independent random strings of length $n$, then $b$ is such a shortest description, but cannot be used as $f$ in Theorem~\ref{th:1}. In this case one can let $f=a\oplus b$ (bit-wise sum modulo~$2$) instead: knowing $f$ and $a$, we reconstruct $b=a\oplus f$,
but $\KS(b|f)\approx n$.

This trick can be generalized to provide an alternative proof for Theorem~\ref{th:1}. For this we use the conditional description theorem from~\cite{conditional-codes}. It says that 
	\begin{quote}
for any two strings $a,b$ of length at most $N$ there exist a string $b'$ such that 

\textbullet\quad$\KS(b|a,b')=O(\log N)$ [$b'$ is a description of $b$ when $a$ is known],  

\textbullet\quad$\KS(b'|b)=O(\log N)$ [$b'$ is simple relative to $b$] and 

\textbullet\quad the length of $b'$ is $\KS(b|a)$ [$b'$ has the minimal possible length for descriptions of $b$ when $a$ is known].
	\end{quote}

To prove Theorem~\ref{th:1}, take this $b'$ and also $a'$ defined in the symmetric way (the short description of $a$ when $b$ is known simple relative to~$a$).  Add trailing zeros or truncate $a'$ to get the string $a''$ that has the same length as $b'$. (Adding zeros is needed when $\KS(a)<\KS(b)$, truncation is needed when $\KS(a)>\KS(b)$.) Then let $f=a''\oplus b'$.

A person who knows $a$ and gets $f$, can compute (with logarithmic additional advice) first $a'$, then $a''$, then $b'$ and then $b$. It is not difficult to check also that $\KS(b|f)=\min\{\KS(a),\KS(b)\}$ with logarithmic precision.

Indeed, 
	\begin{multline*}
\KS(b|f)=\KS(b,f|f)=\KS(b,b',f|f)=\KS(b,a''|f)\ge\\ \ge \KS(b,a'')-\KS(f)\ge\KS(b,a'')-|f|=\KS(b,a'')-\KS(b|a)
	\end{multline*}
with logarithmic precision. The strings $a'$ and $b$ are independent (have logarithmic mutual information), so $b$ and $a''$ (that is a simple function of $a'$) are independent too. Then we get lower bound $\KS(b)-\KS(b|a)+\KS(a'')$ which is equal to $\min\{\KS(a),\KS(b)\}$.
(End of the alternative proof.)  

The advantage of this proof: it provides a message $f$ of polynomial in $N$ length (unlike our original proof, where the message is some function that has domain of exponential size), and, moreover, $f$ has the minimal possible length $\KS(b|a)$.
The result it gives can be stated as follows:

\begin{theorem}
	\label{th:1bis}
For every two strings $a$ and $b$ of length at most $N$ there exists a string $f$ of length $\KS(b|a)$ such that
	$$
\KS(b|f,a)=O(\log N)
	$$
and	
	$$
\KS(b|f)=\min\{\KS(a),\KS(b)\} + O(\log N).    
     $$	
\end{theorem}	

The disadvantage is that this proof does not work for relativized case (Theorem~\ref{th1:relativized}), at least literally. For example, let $a$ and $b$ be independent strings of length $2n$ and let $a=a_1a_2$ and $b=b_1b_2$ be their divisions in two halves. Then let $c=(a_1\oplus a_2\oplus b_1)(a_2\oplus b_1\oplus b_2)$. Then $\KS(a|c)=\KS(a,c|c)=\KS(a,b|c)=2n$, $\KS(b|c)=2n$, but $\KS(b|c,a\oplus b)=0$.

In the next section we provide a different construction of a short message $f$ that has the required properties (contains information about $b$ only for those who know $a$ but not for those who know $c$).

\section{A combinatorial construction of a low complexity\\ description}

We will prove that if $a$ contains enough information (more precisely, if $\KS(a|c)\ge \KS(b|c)+\KS(b|a)+O(\log N)$), then there exists a message $f$ that satisfies the claim of Theorem~\ref{th1:relativized} and has complexity $\KS(b|a)+O(\log N)$. We need the following combinatorial statement. (By $\mathbb{B}^k$ we denote the set of $k$-bit binary strings.)

\subsection*{Combinatorial statement}

\begin{lemma}
   Let $n\ge m$ be two positive integers. There exists a family $\mathcal{F}$ consisting of $2^m\poly(n)$ functions of type $\mathbb{B}^n\to\mathbb{B }^m$ with the following property: for every string $b\in \mathbb{B}^m$ and for every subfamily $\mathcal{F}'$ that contains at least half of the elements of $\mathcal{F}$, there are at most $O(2^m)$ points with the second coordinate $b$ and not covered by the graphs of the functions in $\mathcal{F}'$.
\end{lemma}

\begin{figure}[h]
\begin{center}
\includegraphics{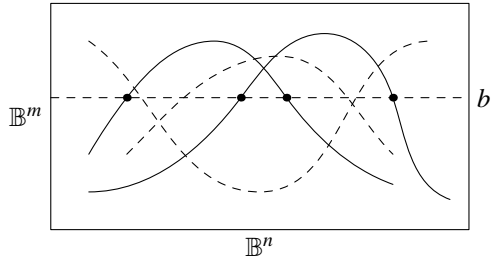}
\end{center}
\caption{Some functions (up to $50\%$) are deleted from $\mathcal{F}$; nevertheless the graphs of the remaining ones cover every horizontal line almost everywhere (except for $O(2^m)$ points).}
\label{fig}
\end{figure}

Formally the property of $\mathcal{F}$ claimed by Lemma (Fig.~\ref{fig}) can be written as follows:
	$$
	\forall b\, \forall \mathcal{F}'\subset\mathcal{F}
	\bigl[\#\mathcal{F}' \ge \frac{1}{2}\#\mathcal{F}\ \Rightarrow\ 
	\#\{ a\in\mathbb{B}^n \mid  f(a)\ne b 
	\text{ for all } f\in\mathcal{F}'\} = O(2^m)
	\bigr].
	$$
	
(Note that the condition $n\ge m$ is in fact redundant: if $n<m$, the claim is trivial since the number of all $a$ is $O(2^m)$.)
	
Before proving the Lemma, let us try to explain informally why it could be relevant. The family $\mathcal{F}$ is a reservoir for messages ($f$ will be a number of some function from $\mathcal{F}$). Most functions from $\mathcal{F}$ (as in any other simple family) have almost no information about $b$; they form $\mathcal{F}'$. If the pair $\langle a, b\rangle$ is covered by the graph of some function $f\in\mathcal{F'}$, then $f$ (i.e., its number) is the required message. If not, $a$ belongs to a small set of exceptions, and its complexity is small, so the condition of the theorem is not satisfied. (See the detailed argument below.)

\textbf{Proof} of the combinatorial lemma. We use probabilistic method and show that for a random family of $2^t$ independent random function the required property holds with positive probability. (The exact value of parameter $t$ will be chosen later.)

Let us upperbound the probability of the event ``random family
	$
\varphi_1,\ldots,\varphi_{2^t}
	$
does not satisfy the required property''. This happens if there exist

\begin{itemize}

\item an element $b\in \mathbb{B}^m$;

\item a set $S\subset \mathbb{B}^n$ that contains $s2^m$ elements (the exact value of the constant $s$ will be chosen later);

\item a set $I\subset\{1,2,\ldots,2^t\}$ that contains half of all indices

\end{itemize}
	
\noindent 
such that
         $$
    \varphi_i(a)\ne b \text{ for every $a\in S$ and every $i\in I$}.
    \eqno(*)  
         $$
To get an upper bound for the probability of this event, note that there are $2^m$ different values of $b$, at most $2^{2^t}$ different values of $I$ and at most $(2^n)^{s2^m}$ different values of $S$. For fixed $b$, $I$,  and $S$ the probability of $(*)$ is
	$$
	\left(1 - \frac{1}{2^m}\right)^{2^{t-1}\cdot s2^m}
	$$
(each of $2^{t-1}$ functions with indices belonging to $I$  has a value different from~$b$ at each point $a\in S$). In total we get an upper bound
     $$
2^m \cdot 2^{2^t} \cdot 2^{ns2^m}\cdot 
\left(1 - \frac{1}{2^m}\right)^{2^{t-1}\cdot s2^m},    
     $$
and we have to show that this product is less than $1$ if the values of the parameters are chosen properly. We can replace $(1-1/2^m)^{2^m}$ by $1/e$ (the difference is negligible with our precision) and rewrite the expression as
	$$
2^{m+2^t}\cdot 2^{ns2^m}\cdot (1/e)^{s2^{t-1}}.
     $$	
The most important terms are those containing $2^t$ and $2^m$ in the exponents (since $2^t,2^m\gg m,n,s$). We want the last small term to overweight the first two. Let us split it into two parts $(1/2)^{s2^t/4}$ and use these parts to compensate the first and the second term. It is enough that
	$$
2^{m+2^t}\cdot (1/e)^{s2^t/4} < 1
	$$
and
	$$
2^{ns2^m}\cdot (1/e)^{s2^t/4} < 1
	$$
at the same time. The first inequality can be made true if the constant $s$ is large enough (note that $m\ll 2^t$). The second inequality (where both exponents can be divided by $s$) is achievable with $2^t=2^m\poly(n)$.~\raisebox{-0.5ex}{\hbox{\small$\Box$}}

\subsection*{Main result}

Now we are ready to give the formal statement and proof:

\begin{theorem}
    \label{th:main}
There exists a constant $d$ such that for any strings $a,b,c$ of length at most $N$ satisfying the inequality
	$$
\KS(a|c)\ge\KS(b|c)+\KS(b|a)+d\log N
	$$
there exists a string $f$ of length at most $\KS(b|a)+d\log N$ such that $\KS(b|a,f)\le d\log N$ and $\KS(b|c,f)\ge\KS(b|c)-d\log N$.
\end{theorem}

Recall the intuition behind this result. The condition of the theorem guarantees that the agent's ``background'' $a$ has enough information not available to the adversary (who knows~$c$); theorem guarantees that there exists a string $f$ that allows the agent to reconstruct $b$ from $a$, has the minimal possible length among all strings with this property and does not provide any information about $b$ if the adversary knows only $c$. (Note that we use the same constant $d$ in all $O(\log N)$ expressions, but this does not matter since increasing $d$ makes the statement only weaker.)

\textbf{Proof}. Using conditional description theorem~\cite{conditional-codes}, we find string $b'$ of length $\KS(b|a)$ such that both complexities $\KS(b|b',a)$ and $\KS(b'|b)$ are $O(\log N)$. Then we apply the combinatorial lemma with $n$ equal to the length of $a$ and $m$ equal to the length of $b'$, i.e., to $\KS(b|a)$. The lemma provides a family $\mathcal{F}$, and we may assume without loss of generality that the complexity of $\mathcal{F}$ is $O(\log N)$ (for given $m$ and $n$, take the first family with the required properties in some fixed ordering).

Most functions in $\mathcal{F}$ (as well as most objects in any simple set) do not have much information about $b$ when $c$ is known, i.e., the difference $\KS(b|c)-\KS(b|f,c)$ is small for most $f\in\mathcal{F}$. Indeed, with logarithmic precision this difference can be rewritten as $\KS(f|c)-\KS(f|b,c)$ (recall the formula for pair and conditional complexities), and the average value of both terms in the last expression is $m+O(\log N)$, the difference is of order $O(\log N)$ and we can use the Chebyshev inequality.

Let $\mathcal{F}'$ be functions from this majority. The lemma guarantees that the graphs of those functions cover all pairs $\langle a',b'\rangle$ for all strings $a'$ of length $n$ except for $O(2^m)$ ``bad'' values of $a'$, and it remains to show that the given string $a$ is not ``bad''. It is because
	$$
\KS(a'|c)<\KS(b|c)+\KS(b|a)+O(\log N)
	$$
for all ``bad'' $a'$. Indeed, knowing $b$, $c$ and $\KS(b|c)$ (the latter contains $O(\log N)$ bits and can be ignored with logarithmic precision), we can enumerate all functions $f$ that do not belong to $\mathcal{F'}$ (=functions that make complexity of $b$ with condition $c$ smaller), and therefore we can enumerate all $O(2^m)$ ``bad'' values. (Note also that $b'$ can also be obtained from $b$ with a logarithmic advice.) So the complexity of the ``bad'' values (for known $b$ and $c$) is at most $m+O(\log N)$:
	$$
\KS(a'|b,c)\le \KS(b|a)+O(\log N)
	$$
for all ``bad'' $a'$, therefore
	$$
\KS(a'|c)\le \KS(a'|b,c)+\KS(b|c)+O(\log N) \le \KS(b|a)+\KS(b|c)+O(\log N)
	$$
as we claimed.~\raisebox{-0.5ex}{\hbox{\small$\Box$}}

\section{Negative result and open questions}

Theorem~\ref{th:main} makes an assumption that looks artificial at first: for example, if $a, b, c$ are pairwise independent, we require $\KS(a)$ to be twice as big as $\KS(b)$, and it is intuitively unclear why the amount of the background information should be twice as big as the message we want to transmit (inequality $\KS(a)>\KS(b)$ looks more natural). In this section we show that this condition, even if looking artificial, is important: without it, all the strings $f$ that satisfy the claim of Theorem~\ref{th1:relativized} may have exponentially large length. The exact statement (see Theorem~\ref{th:negative} below) and its proof are rather technical, so let us start with a simplified example, where, unfortunately, we get a string $c$ of large complexity. Then we explain the more advanced example that does not have this problem.

Let us construct three strings $a,b,c$ with the following properties: every reasonably long program $f$ (of polynomial or subexponential length) that maps $a$ to $b$ can be used to simplify the transformation of $c$ into $b$. In our example the string $a$ has complexity $1.3n$, the string $b$ has complexity $n$, and they are mutually independent (have logarithmic mutual information). (The coefficient $1.3$ is chosen arbitrarily; it is important that $1.3$ is greater than~$1$ and less than~$2$. The complexity of $b$ when $c$ is known will be about~$n$, so using $c$ as a condition does not make $b$ simpler. But if we add to $c$ any program $f$ that maps $a$ to $b$, it becomes possible to obtain $a$ using only $0.3n$ bits of advice: the conditional complexity decreases from $\KS(b|c)\approx n$ to $\KS(b|f,c)\approx 0.3n$.

The main idea of this example can be explained as follows: the string $c$ itself encodes a function that maps $a$ to $b$ (but still $c$ without $a$ has no information about $b$). Assume that some program $f$ that maps $a$ to $b$ is given. Why does it help to describe $b$ if $c$ is known in addition to $f$? We know that both $f$ and $c$ map $a$ to $b$, so $a$ is one of the solution of the equation $f(x)=c(x)$. If this equation has not too many solutions, we can describe $a$ (and therefore $b$) by specifying the ordinal number of $a$ in the enumeration of all solutions. (Note that $f$ may be not everywhere defined, but this does not matter.) In this way we get a conditional description of $b$ (for known $c$ and $f$) that may have small length compared to $\KS(b)$ (and $\KS(b)$ will be close to $\KS(b|c)$; we promised that $c$ itself has no information about $b$).

How do we get $a$, $b$, and $c$ with these properties? We get such a triple with high probability if $a$ and $b$ are independently taken at random among strings of length $1.3n$ and $n$ respectively, and $c$ is a random function whose graph contains pair $\langle a,b\rangle$. The same distribution on $a,b,c$ can be described in a different way: we take a random function $c$ and then a random element $\langle a,b\rangle$ of its graph.

With high probability we get strings $a$ and $b$ with the required complexities $1.3n$ and $n$ and small mutual information. We can also show that $\KS(b|c)$ is close to $n$ with high probability. Indeed, for a typical function $c$ of type $\mathbb{B}^{1.3n}\to\mathbb{B}^n$ most of its values have preimage of size $2^{0.3n}$, and therefore the second component of a random element of its graph has almost uniform distribution, so most of the values of $c$ have high complexity even with condition~$c$.	

Now let $f$ be some program that maps $a$ and $b$ and has not very high complexity (much less than what Theorem~\ref{th1:relativized} gives). How many solution has the equation $f(x)=c(x)$? Typically (for a given~$f$ and a random~$c$) we have about $2^{0.3n}$ solutions (for each $x$ the probability of $f(x)=c(x)$ equals $2^{-n}$, and there are $2^{1.3n}$ points~$x$); here we assume that $f$ is total, but if it is not, we get even less solutions. For a fixed $f$ and a random $c$, it is very unlikely that the number of solutions is significantly greater than $2^{0.3n}$. In other words, Hamming ball of the corresponding radius around $f$ has a negligible probability. If the number of these balls (i.e., the number of programs $f$ we consider) is not too large, the union of these events also has small probability,  so a randomly chosen $c$ will be outside these balls. This means that for all programs $f$ with bounded complexity the equation $f(x)=c(x)$ has at most $2^{0.3n}$ solutions (or slightly more) and the complexities $\KS(a|f,c)$ and $\KS(b|f,c)$ are (almost) bounded by $0.3n$ as we promised.

We do not provide details of this argument since we want to prove a stronger (and more complicated) results. Namely, we want to find a function $c$ that has not very high complexity (and the argument explained gives $c$ that can have exponential complexity): the complexity of $c$ should exceed the complexity of programs $f$ (that it opposes) by $\KS(b)$. (If we allow more programs, we need more freedom for $c$.)

The idea of the construction remains the same: we select a random point on the graph of a random function. However, now the function is a random element of some family $\mathcal{C}$ of functions. We formulate some combinatorial properties of $\mathcal{C}$. Then we prove (by a probabilistic argument) that there exists a family with these properties and conclude that there exists a simple family with these properties (the first family found by exhaustive search). Finally, we prove that for most pairs $\langle a,b\rangle$ there exists a function $c$ in the family that satisfies our requirements. (So we prove even a bit stronger statement: instead of existence of a triple $a,b,c$ we prove that for most $a$ and $b$ there exists $c$.)
The size of the family $\mathcal{C}$ provides a bound for the complexity of $c$ (since every element of $\mathcal{C}$ is determined by its index).

Let us formulate the required combinatorial statement starting with some definitions. Fix some sets $A$ and $B$. We say that some family $\mathcal{F}$ of functions $A\to B$ \emph{rejects} a function $c\colon A\to B$ if there exists $f\in\mathcal{F}$ such that the cardinality of the set $\{a\colon c(a)=f(a)\}$ exceeds $4\#A/\#B$ (note that the ``expected'' cardinality is $\#A/\#B$). Let $\mathfrak{H}$ be a mapping defined on $B$; for every $b\in B$ the value $\mathfrak{H}(b)$ is a family of functions of type $A\to B$ (i.e., $\mathfrak{H}(b)\subset B^A$ for every $b\in B$). We say that a function $c$ \emph{covers} the pair $\langle a,b\rangle\in A\times B$ (for given $\mathfrak{H}$ and $\mathcal{F}$) if (1)~$c(a)=b$; (2)~the function $c$ is not rejected by $\mathcal{F}$ and (3)~$c\notin\mathfrak{H}(b)$. 

\begin{lemma}\label{combinatorial-negative}
Assume that $\#B\ge 2$ and $\#A\ge 16\#B$. Assume that two numbers $\varepsilon\ge4\#B/\#A$ and $\phi\le 2^{\#A/(4\#B)}$ are fixed. There exists a family $\mathcal{C}$ of functions $A\to B$ of cardinality 
	$$ 
\max\left\{\frac{20\#B}{\varepsilon},\: \frac{6\Phi\log_2(\#B)}{\varepsilon},\: 
       6\Phi\cdot\#B\cdot\log_2(\#B)\right\}
	$$
with the following property: for every family $\mathcal{F}$ of size at most $\Phi$ and for every mapping $\mathfrak{H}$ such that $\#(\mathfrak{H}(b))\le(1/4)\#\mathcal{C}$ for every $b\in B$, at most $\varepsilon$-fraction of all pairs $\langle a,b\rangle$ are not covered by any $c\in \mathcal{C}$ \textup(for these $\mathcal{F}$ and $\mathfrak{H}$\textup).
\end{lemma}

The statement of this lemma can be written as follows (we omit conditions for cardinalities of $\mathcal{C}$, $\mathcal{F}$ and $\mathfrak{H}(b)$):
        \begin{multline*}
\exists \mathcal{C}
\ \forall \mathcal{F},\mathfrak{H}\\ 
\biggl|\biggl\{\langle a,b\rangle\colon
\forall c\: 
\biggl[ (c(a)=b) \Rightarrow \bigl[(c\in \mathfrak{H}(b))\ \lor\ 
            (\exists f\in \mathcal{F}\:  
                  \#\{x:f(x)=c(x)\}\ge{\textstyle\frac{4\#A}{\#B}}
)\bigr]\biggr]
\biggr\}\biggr|\le\\ \le \varepsilon\cdot\#A\cdot\#B\,.
        \end{multline*}

Let us explain informally the meaning of this lemma (how it is used in the sequel). We may assume without loss of generality that the family $\mathcal{C}$ is simple (looking for the first family with the required properties in some ordering). Let $\mathcal{F}$ be the family of all functions that have simple programs (or their extensions, if the functions are partial). Let $\mathfrak{H}(b)$ be the set of all functions that are simple when $b$ is known (having small conditional complexity with condition $b$).  For a pair $\langle a,b\rangle$ that does not belong to the ``bad'' $\varepsilon$-fraction, there exists a function $c\in\mathcal{C}$ that covers $\langle a,b\rangle$. This function (or, better to say, its index in $\mathcal{C}$) is a counterexample we are looking for. Indeed, if the eavesdropper knows $c$ and gets a simple program $f$ mapping $a$ to $b$, the complexity of $b$ for her decreases. Indeed, it is enough to specify the ordinal number of $a$ in the enumeration of all solutions of the equation $f(x)=c(x)$, and the eavesdropper can reconstruct $a$ (and therefore $b$, since $f(a)=b$). On the other hand, the choice of $\mathfrak{H}$ guarantees that $c$ and $b$ are independent (i.e., $c$ has maximal possible complexity even if $b$ is known). The details of these argument will be explained later, after we prove the lemma.

\textbf{Proof of the lemma}.
Using a probabilistic argument, let us consider a random family $\mathcal{C}$ of the size mentioned. We assume that $\mathcal{C}$ is indexed by integers in range $1$\ldots$\#\mathcal{C}$, and for every index $i$ and every point $a\in A$ the value of $i$th function on $a$ is an independent random variable uniformly distributed over~$B$. Then we prove that the probability of the event ``$\mathcal{C}$ is bad'' (i.e., does not have the required property) is strictly less than~$1$.

For this we get an upper bound for the probability of the event ``$\mathcal{C}$ does not have the required property'' with respect to a fixed family~$\mathcal{F}$ (and then multiply it by the number of different families $\mathcal{F}$). So let us assume $\mathcal{F}$ is fixed. Things are ``good'' if for every mapping $b\mapsto\mathfrak{H}(b)$ (with our restrictions: all $\mathfrak{H}(b)$ have cardinality at most $(1/4)\#\mathcal{C}$) for $\varepsilon$-almost all pairs $\langle a,b\rangle$ there is a function $c\in\mathcal{C}$ that is not rejected by $\mathcal{F}$ and is not in $\mathfrak{H}(b)$ such that $c(a)=b$.

Note that the definition of rejection does not refer to $\mathcal{C}$: the set of rejected function is determined by $\mathcal{F}$ alone. For a given $\mathcal{F}$ there are two possibilities: (1)~many functions are rejected (we choose $(1/4)\#\mathcal{C}$ as a threshold) or (2)~not many functions are rejected. In the latter case we may add rejected functions to all $\mathfrak{H}(b)$ (for all $b$), and the size of all $\mathfrak{H}(b)$ remains bounded by $(1/2)\#\mathcal{C}$.

In other term, for a fixed $\mathcal{F}$ the ``bad'' event is covered by the union of the following two events:
	\begin{enumerate}
\item $\mathcal{F}$ rejects at least $1/4$ of all functions in $\mathcal{C}$;

\item there exists a mapping $b\mapsto\mathfrak{H}(b)$ where all sets $\mathfrak{H}(b)$ have cardinality at most $(1/2)\#\mathcal{C}$ such that the fraction of pairs $\langle a,b\rangle\in A\times B$ that do not belong to any function $c\in\mathcal{C}\setminus\mathfrak{H}(b)$ exceeds~$\varepsilon$.
	\end{enumerate}
	
What we need is the following: the sum of the probabilities of these two events multiplied by the number of possibilities for $\mathcal{F}$ is less than~$1$. To show this, we prove that each of these two probabilities is less than $1/2$ divided by $(\# B^{\# A})^{\Phi}$ (this expression is an upper bound for the number of different families $\mathcal{F}\subset B^A$ of size $\Phi$).

The first event can be rewritten as follows: \emph{there exists a subfamily $\mathcal{C}'\subset\mathcal{C}$ of size $\#\mathcal{C}/4$ such that for all $c\in\mathcal{C}'$ there exists $A'\subset A$ of size $4\#A/\#B$ and a function $f\in\mathcal{F}$ such that $f(a)=c(a)$ for all $a\in A'$.}

The number of possibilities for $\mathcal{C}'$ does not exceed $2^{\#\mathcal{C}}$, the number of all subsets. For a fixed $\mathcal{C}'$ (or, better to say, for a fixed set of indices) the functions with these indices are chosen independently. So we can estimate the probability of the bad event for one index and then use independence.  To get an upper bound for the number of possibilities for $A'$ let us note that the number of $r$-element subsets of a $q$-element set, $\binom{q}{r}$, does not exceed  $q^r/r!\le {q^r}/{((r/3)^r)}=(3q/r)^r$. For $q=\#A$ and $r=4\#A/\#B$ we get the bound $(3\#B/4)^{4\#A/\#B}$.

Therefore, the probability of the first event does not exceed
$$
   2^{\#\mathcal{C}}\left(
     \Phi
     \left(\frac{3\#B}{4}\right)^{4\#A/\#B}
     \left(\frac{1}{\#B}\right)^{4\#A/\#B}
   \right)^{\#\mathcal{C}/4}=
   \left(2\Phi^{1/4}
     \left(\frac{3}{4}\right)^{\#A/\#B}
   \right)^{\#\mathcal{C}}\,.
$$
Multiplied by $(\#B)^{\#A\cdot\Phi}$ (the number of possibilities for~$\mathcal{F}$), this probability is less than $1/2$, since $\#B\ge 2$, $\#A\ge 16\#B$,  $\Phi\le 2^{\#A/4\#B}$, and $\#\mathcal{C}\ge 6\Phi\cdot\#B\cdot\log_2(\#B)$ (according to lemma's assumptions). Indeed, the last inequality implies that $\#\mathcal{C}\ge 12$ if $\Phi\ge 1$ (for empty $\mathcal{F}$ the statement is trivial). Since $\#B\ge 2$, we conclude that $1+\#\mathcal{C}\le 13\#\mathcal{C}/12$. Then $1\le\#A/(16\#B)$ implies that $1+\#\mathcal{C}\le (13/192)(\#A\cdot\#C/\#B)$. The condition $\log_2\Phi\le \#A/4\#B$ implies that $(\#\mathcal{C}/4)\log_2\Phi \le (1/16)(\#A\cdot\#\mathcal{C}/\#B)$. Finally, the inequality $\#\mathcal{C}\ge 6\Phi\cdot\#B\cdot\log_2\#B$ implies that $\#A\cdot\Phi\log_2\#B\le (1/6)(\#A\cdot\#\mathcal{C}/\#B)$. Adding these inequalities (note that $19/64 < 1/3 < \log_2(4/3)$ and taking the exponent (with base $2$) of both sides, we get the required inequality (after appropriate grouping of the factors).

Now let us consider the second event (recall that it depends on $\mathcal{F}$ which is fixed): \emph{there exist a mapping $b\mapsto\mathfrak{H}(b)$ such that every $\mathfrak{H}(b)$ has cardinality at most $\#\mathcal{C}/2$ and a subset $U\subset A\times B$ of size $\varepsilon\cdot\#A\cdot\#B$ such that for every pair $\langle a,b\rangle\in U$ and for every function $c\in\mathcal{C}\setminus\mathfrak{H}(b)$ we have $c(a)\ne b$.}

In the sequel we assume that $\mathfrak{H}(b)$ is not a set of functions, but a set of their indices (numbers in $1$\ldots$\#\mathcal{C}$ range); this does not change the event in question.

To estimate the probability of the second event, let us fix not only $\mathcal{F}$ but also $\mathfrak{H}$ and $U$. The corresponding event can be described as the intersection (taken over all pairs $\langle a,b\rangle$ and over all $i\notin\mathfrak{H}(b)$) of the events $c[i](a)\ne b$ (``the $i$th function does not map $a$ to $b$'').
The probability bound would be simple if all these events were independent; in this case the probability would be $(1-1/\#B)^d$, where $d$ is the number of all triples $\langle i,a,b\rangle$, i.e., $\varepsilon\cdot \#A\cdot\#B\cdot\#\mathcal{C}/2$ (i.e., $d$ is the product of the number of pairs $\langle a,b\rangle\in U$ and the number of possible values of $i$ for given $b$).

Unfortunately, these events are independent only for different $a$ (or different $i$); the events $c[i](a)\ne b_1$ and $c[i](a)\ne b_2$ are dependent. However, the dependence works in the ``helpful'' direction: the condition $c[i](a)\ne b_1$ only increases the probability of the event $c[i](a)\ne b_2$ (the denominator in $1/\#B$ decreases by $1$). The same is true for several conditions. 

Formally speaking, we may group the events with common $a$ and $i$ and then use the inequality $(1-k/\#B)\le (1-1/\#B)^k$, where $k$ is the number of events in a group.

In this way we get an upper bound for the probability of failure: for fixed $\mathcal{F}$, $\mathfrak{H}$ and $U$, it does not exceed 
	$$
\left(1-\frac{1}{\#B}\right)^{\varepsilon\cdot\#A\cdot\#B\cdot\#\mathcal{C}/2}\le
2^{-\varepsilon\cdot\#A\cdot\#\mathcal{C}/2}\,.
	$$
This expression is then multiplied by the number of possibilities for $U$ (that does not exceed $2^{\#A\cdot\#B}$), for $\mathfrak{H}$ (that does not exceed $(2^{\#\mathcal{C}})^{\#B}$) and for $\mathcal{F}$. In total, we get
	$$
2^{-\varepsilon\cdot\#A\cdot\#\mathcal{C}/2}\cdot
2^{\#A\cdot\#B}\cdot
2^{\#\mathcal{C}\cdot\#B}\cdot
(\#B)^{\#A|\cdot\Phi}\,.
	$$
It is easy to check that this expression is less than $1/2$ if $\#B\ge 2$, $\varepsilon\ge 4\#B/\#A$, $\#\mathcal{C}\ge 20\#B/\varepsilon$,  and $\#\mathcal{C}\ge (6\Phi\log_2\#B)/\varepsilon$. Indeed, we have $1+\#A\cdot\#B\le 3\cdot\#A\cdot\#B/2$ if $A$ is not empty and $\#B\ge 2$. Therefore, $\#\mathcal{C}\ge 20\cdot\#B/\varepsilon$ implies $1+\#A\cdot\#B\le (3/40)\varepsilon\cdot\#A\cdot\#\mathcal{C}$.
Also $\varepsilon\ge 4\#B/\#A$ implies
$\#\mathcal{C}\cdot\#B\le (1/4)\varepsilon \cdot\#A\cdot\#\mathcal{C}$.
Finally, $\#\mathcal{C}\ge (6\Phi\log_2\#B)/\varepsilon$ implies
$\#A\cdot\Phi\cdot\log_2(\#B)\le (1/6)\varepsilon\cdot\#A\cdot\#\mathcal{C}$. Adding these inequalities, noting that $59/120 < 1/2$ and then taking the exponents (with base $2$), we get the required bound after regrouping the factors.

Lemma is proven.

Now we use this lemma to prove the promised negative result. Let $\alpha>0$ be some constant. Let $m,n,l$ be positive integers such that $n\ge 1$, $m\ge n+4$, $m-\alpha\log_2 m\ge n+2$, and $l+1+\log_2 (l+1)\le 2^{m-n-2}$. Let $N=\max\{m,l\}$.

\begin{theorem}\label{th:negative}
Let $a$ be a string of length $m$ and let $b$ be a string of length $n$ such that
	$$
m+n-\KS^{\mathbf{0}'}(a,b)<\alpha\log_2 m.
	$$
Then there exists a string $c$ of complexity $n+l+O(\log N)$ such that 
	\begin{itemize}
\item $\KS(c|b)=\KS(c) + O(\log N)$;
\item $\KS(b|a,c)=O(\log N)$;
\item for every $f$ such that  $\KS(f)\le l-\KS(b|a,f)$ 
      we have $\KS(b|c,f)\le m-n + \KS(b|a,f) +O(\log N)$.
	\end{itemize}
\textup(The constant hidden in $O(\cdot)$ depends on $\alpha$ but not on
$m$, $n$, $l$.\textup)
\end{theorem}

Before proving this theorem, let us explain why it shows the importance of the condition in theorem~\ref{th:main}. The equation $\KS(c|b)=\KS(c) + O(\log N)$ shows that the strings $b$ and $c$ are independent and $\KS(b|c)=\KS(b)=n$ with $O(\log N)$-precision. Since $\KS(b|a,c)=O(\log N)$, we have $\KS(a|c)\ge\KS(b|c)-\KS(b|a,c)=n$ (with the same $O(\log N)$-precision). Note also that $\KS(b|a)= n$ (with $O(\log m)$-precision). Therefore, if $\KS(b|a,f) = O(\log N)$ for some string $f$ of length not exceeding $l$, then
	$$
\KS(b|c,f) < \min\{\KS(a|c),\KS(b|c)\} + O(\log N)
	$$ 
when $m - n < n + O(\log N)$, i.e., when $\KS(a) < \KS(b|c) + \KS(b|a)$.

\textbf{Proof}. Let $A$ be the set of all $m$-bit strings, and let $B$ be the set of all $n$-bit strings. Let $\varepsilon=1/m^\alpha$ and $\Phi=2^l(l+1)$. Our assumptions about $n,m,l$ guarantee that $A$, $B$, $\varepsilon$ and $\Phi$ satisfy the conditions of the lemma. Therefore there is a family $\mathcal{C}$ with the properties described in the statement of the lemma. As we have said, we may assume without loss of generality that $\mathcal{C}$ is simple, and in this case the complexity of every element of $\mathcal{C}$ does not exceed $\log_2\#\mathcal{C}$ plus $O(\log N)$, i.e., does not exceed $n+l+O(\log N)$.

Now let $\mathfrak{H}(b)$ be the set $\{c\in \mathcal{C}\colon \KS(c|b) < \log_2(\#\mathcal{C})-2\}$; then $\#\mathfrak{H}(b)\le \#\mathcal{C}/4$ for every~$b$.

Now the family $\mathcal{F}$ is constructed as follows. It contains $\Phi$ functions numbered by integers in $1$\ldots$\Phi$ range. We enumerate all triples $\langle a,b,f\rangle$, 
where $a\in A$, $b\in B$ and $f$ is a $l$-bit string such that $\KS(f) + \KS(b|a,f)\le l$. Some indices (numbers) have labels that are $l$-bit strings. When a new triple  $\langle a,b,f\rangle$ appears, we first try to add $\langle a,b\rangle$ to one of the functions whose index already has label $f$. If this is not possible (all functions that have label $f$ are already defined at $a$ and have values not equal to $b$), we take a fresh index (that has no label), assign label $f$ to it and let the corresponding function map $a$ to $b$. A free index does exist since each $f$ occupies at most $2^{l-\KS(f)+1}$ indices (if some $f$ needs more, then for some $a$ all $2^{l-\KS(f)+1}$ functions are defined and have different values, so we have enumerated already more than $2^{l-\KS(f)+1}$ different elements $b$ such that 
$\KS(b|a,f)\le l-\KS(f)$;~a contradiction), and all $f$ in total require at most $\sum_{\KS(f)\le l}2^{l-\KS(f)+1}=\sum_{k=0}^l\sum_{\KS(f)=k}2^{l-k+1}=\Phi$ indices. After all the triples with these properties are enumerated, we extend our functions to total ones (arbitrarily).

Consider the set of pairs $\langle a,b\rangle$ that are not covered by $\mathcal{C}$ (for given $\mathcal{F}$ and $\mathfrak{H}$). The cardinality of this set does not exceed $\varepsilon2^{m+n}$. On the other hand, $\mathcal{F}$ and $\mathfrak{H}$ can be computed using $\mathbf{0}'$-oracle, and after that the set of non-covered pairs can be enumerated, therefore $\KS^{\mathbf{0}'}(a,b)\le m+n-\alpha\log_2 m$ for every non-covered pair $\langle a,b\rangle$.

Therefore for every $a$ and $b$ such that $m+n-\KS^{\mathbf{0}'}(a,b)<\alpha\log_2 m$ there exists $c\in\mathcal{C}$ such that $c(a)=b$, $c\notin \mathfrak{H}(b)$, and for every 
$f\in \mathcal{F}$ the equation $c(x)=f(x)$ has at most $2^{m-n+2}$ solutions.

Since $c(a)=b$, we have $\KS(b|a,c)=O(\log N)$.

Since $c\notin \mathfrak{H}(b)$, we have $\KS(c|b) \ge \log_2(\#\mathcal{C})-2$, i.e.,  $\KS(c)=\KS(c|b)+O(\log N)$.

Finally we have to estimate $\KS(b|c,f)$ for strings $f$ such that $\KS(f)\le l-\KS(b|a,f)$. Knowing $f$, we enumerate functions in $\mathcal{F}$ that have label~$f$. One of them, say, $\tilde f$, goes through $\langle a,b\rangle$ (i.e., $\tilde{f}(a)=b$). To specify this functions, we need at most $\KS(b|a,f)+O(\log N)$ additional bits. Knowing $\tilde{f}$ and $c$ we may enumerate all $x$ such that $c(x)=\tilde{f}(x)$. (More precisely, we specify the index of $\tilde{f}$ in $\mathcal{F}$, not the $\tilde{f}$ itself. However, to enumerate the solutions of the equation $c(x)=\tilde{f}(x)$ it is enough to enumerate pairs $\langle x,y\rangle$ such that $y=\tilde{f}(x)$ by replaying the construction of $\mathcal{F}$.) This set contains $a$ and has cardinality at most $2^{m-n+2}$, so we can specify $a$ using $m-n+2$ additional bits. Altogether, $\KS(b|c,f)\le \KS(a|c,f) + O(\log N)\le \KS(b|a,f) + m - n + O(\log N)$, as we claimed.

Theorem~\ref{th:negative} is proven.

\bigskip
\textbf{Open questions}

\begin{enumerate}
\item Is it possible to strengthen theorem~\ref{th:negative} and have $c$ of complexity at most $n+O(\log N)$ instead of $n+l+O(\log N)$?  (An.A.~Muchnik in his talk claimed that this can be done by a more complicated combinatorial argument, which was not explained in the talk.)

\item Theorem~\ref{th:negative} shows that if $a$ is only slightly more complex than $b$, then for some $c$ short messages do not work. On the other hand, the alternative proof of theorem~\ref{th:1} works for empty~$c$. What can be said about other $c$? What are the conditions that make short messages possible?

\item What can be said about the possible complexities $\KS(f|b)$, $\KS(f|a,b)$, and $\KS(f|a,b,c)$ if $f$ is a message with the required properties?

\end{enumerate}

\end{document}


\title{Колмогоровская сложность и криптография}
\author{Ан.\,А.\,Мучник\thanks{%
        Результаты этой статьи были доложены Ан.\,А.\,Мучником
        (1958--2007) на колмогоровском семинаре кафедры
        математической логики и теории алгоритмов МГУ
        (заседания 11~марта и 8~апреля 2003~года), но не
        были тогда опубликованы; их формулировки
        составили содержание доклада Ан.\,А.\,Мучника и
        А.\,Л.\,Семјнова на семинаре в Dagstuhl (Seminar~03181,
        27.04.2003--03.05.2003). Текст подготовили А.\,Чернов и
        А.\,Шень в 2008--2009~годах.}
}

\date{}

\maketitle

\begin{abstract}
С точки зрения колмогоровской сложности рассматриваются задачи о
построении сообщений, которые содержат разное количество
информации о заданном объекте в зависимости от того, какой
дополнительной информацией располагает получатель.

Предположим, что получатель знает слово $a$, и мы хотим сообщить
получателю информацию о некотором слове $b$, при этом таким
образом, чтобы наше сообщение само по себе (без $a$) не
позволяло восстановить $b$. Оказывается, что это возможно (если
слова $a$ и $b$ не слишком просты).

Далее рассматриваются более сложные родственные вопросы: что
будет, если ``противник'' знает некоторую информацию $c$?
насколько длинным должно быть сообщение? Мы уточняем эти
вопросы, указываем условия, при которых сообщение может иметь
полиномиальную длину, и показываем, что они существенны.
\end{abstract}

\section{\texorpdfstring{Неинформативное условное описание}
{Non-informative conditional descriptions}}

В этом разделе мы для данных слов $a$ и $b$ (при некоторых
ограничениях) построим слово $f$, которое позволяет получить $b$
из $a$, но само по себе (без $a$) не содержит информации
относительно~$b$, а также рассмотрим некоторые обобщения этой
задачи.

\subsection*{Равномерная и неравномерная сложность}

Начнјм с некоторых общих замечаний, мотивирующих постановку
задачи с точки зрения колмогоровской сложности условного
описания. Пусть $X$ --- некоторое множество (двоичных) слов, а
$y$ --- слово. Тогда можно определить $\KS(X\to y)$ как
минимальную длину программы, которая дајт $y$ на любом входе
$x$ из $X$. (Как обычно, мы фиксируем некоторый оптимальный
способ записи программ; можно также вместо минимальной длины
говорить о минимальной сложности.) Очевидно, что
    $$
\KS(X\to y) \ge \max_{x\in X} \KS(y|x)
    $$
(если программа $p$ годится для всех $x\in X$, то она годится
для любого из них), но обратное верно не всегда и левая часть
может быть существенно больше правой. Можно сказать, что
``равномерная'' сложность задачи $X\to y$ (левая часть) может
быть существенно больше ``неравномерной''.

Чтобы убедиться в этом, возьмјм в качестве $y$ несжимаемое слово
длины $n$, а в качестве $X$ возьмјм множество всех слов $x$, для
которых $\KS(y|x)<n/2$. Тогда правая часть по построению не
превосходит $n/2$. Покажем, что левая не меньше $n-O(\log n)$. В
самом деле, пусть $p$ --- некоторая программа, которая дајт $y$
на любом входе $x$, для которого $\KS(y|x)<n/2$. Среди таких
входов есть слова сложности $n/2+O(\log n)$, и вместе с $p$ их
достаточно для получения $y$, поэтому $\KS(y|p)\le n/2 + O(\log n)$.
Значит, существует слово $e$ длины $O(\log n)$, для
которого $\KS(y|\langle p,e\rangle)<n/2$. По предположению
$p(\langle p,e\rangle)=y$ и потому сложность $p$ не меньше
$n-O(\log n)$.

\textbf{Замечание}. Множество $X$ можно сделать конечным,
ограничившись, скажем, словами длины не более $2n$.

\medskip

\subsection*{Сложность задачи $(a\to b)\to b$}

Построенный пример показывает, что равномерная и неравномерная
сложности могут сильно отличаться. В следующем примере это не
так, но доказательство их совпадения требует дополнительных
усилий.

Пусть~$a,b$ --- двоичные слова. Определим $(a\to b)$ как
множество всех программ, которые на входе $a$ дают $b$.
Известно~\cite{sv}, что
      $$
\KS((a\to b)\to b)=\min\{\KS(a),\KS(b)\}+O(\log N)
      $$
для любых слов $a,b$ длины не более $N$. Оказывается, что это
утверждение можно усилить, заменив равномерную сложность на
неравномерную.

\begin{thm}
        \label{basic}
Для любых слов $a$ и $b$ длины не больше $N$
найдјтся программа $f$, переводящая $a$ в $b$,
для которой
        $$
\KS(b|f)=\min\{\KS(a),\KS(b)\} + O(\log N).
        $$
\end{thm}

\begin{proof}
В одну сторону ($\le$) неравенство очевидно для любой программы
$f$, переводящей $a$ в $b$. В самом деле, имея такую программу и
любое из слов $a$ и $b$, можно восстановить $b$.

Докажем, что для некоторой функции $f$ верно и обратное
неравенство. Пусть $m$~--- максимальная из сложностей $\KS(b|f)$
для всех функций, переводящих $a$ в $b$. (При этом мы
ограничиваемся конечными функциями, определјнными на множестве
всех слов длины не более $N$ со значениями в том же множестве,
так что они являются конечными объектами и условная сложность
естественно определяется. Заметим также, что с логарифмической
точностью нет разницы между функциями и их программами.)
Нам надо показать, что сложность хотя
бы одного из слов $a$ и $b$ не превосходит $m+O(\log N)$.

Рассмотрим множество $S$ пар $\langle a',b'\rangle$ слов длины
не больше $N$, для которых выполнено такое свойство:
\emph{$\KS(b'|f)\le m$ для всякой функции $f$, определјнной на
всех словах длины не более $N$ и принимающей значения в том же
множестве, для которой $f(a')=b'$}. По построению пара $\langle
a,b\rangle$ входит в $S$.

Множество $S$ можно перечислять, зная $N$ и $m$. В ходе этого
перечисления будем браковать пары, у которых абсцисса или
ордината встречается не в первый раз (среди незабракованных); от
$S$ останется некоторое подмножество $\tilde S$, которое
является графиком некоторого взаимно однозначного соответствия.
Пара $\langle a,b\rangle$ уже не обязана принадлежать $\tilde
S$, но в $\tilde S$ есть либо пара с абсциссой $a$, либо пара с
ординатой $b$ (иначе мы бы не выбросили $\langle a,b\rangle$).

Поэтому достаточно показать, что $\tilde S$ содержит не более
$O(2^m)$ элементов (тогда порядковый номер указанной пары в
перечислении множества $\tilde S$ содержит $m$ битов и позволяет
восстановить либо~$a$, либо~$b$).

В самом деле, множество $\tilde S$ можно расширить до графика
некоторой функции $g$. Если $\langle a',b'\rangle\in \tilde S$,
то $g(a')=b'$ и потому $\KS(b'|g)\le m$ по построению ($\tilde
S$ является подмножеством $S$). Значит, в $\tilde S$ имеется не
более $O(2^m)$ различных $b'$, а по построению их столько же,
сколько пар в $\tilde S$.
\end{proof}

\subsection*{Криптографическая интерпретация}

Утверждение предыдущего пункта имеет следующую интерпретацию. Мы
хотим передать информацию о слове~$b$ некоторому человеку, уже
знающему слово~$a$, послав ему некоторое сообщение $f$. (Вместе
с~$a$ это сообщение позволит легко восстановить слово~$b$.) При
этом мы хотим, чтобы для непосвящјнных, то есть людей, не
знающих слова~$a$, сообщение $f$ несло минимально возможную
информацию о~$b$, то есть чтобы сложность $\KS(b|f)$ была
максимально возможной. Как мы уже видели, эта сложность не может
быть больше $\KS(b)$ и не может быть больше $\KS(a)$ (поскольку
вместе с~$a$ сообщение~$f$ позволяет легко восстановить~$b$).
Теорема~\ref{basic} показывает, что эта граница
($\min\{\KS(a),\KS(b)\}$) действительно достигается.

Можно рассмотреть релятивизованный вариант этой задачи, когда
непосвящјнный знает некоторое третье слово $c$. Наше рассуждение
(соответственно релятивизованное) позволяет найти функцию $f$,
переводящую $a$ в $b$, для которой
        $$
\KS(b|f,c) = \min\{\KS(a|c),\KS(b|c)\};
        $$
эта функция содержит минимально возможную информацию о $b$ (с
точки зрения знающих $c$). Формально говоря, верно следующее
утверждение (и его доказательство повторяет доказательство
теоремы~\ref{basic}):

\begin{thm}
        \label{relativized}
Пусть $a,b,c$~--- слова длины не более $N$. Тогда найдјтся слово
$f$, для которого:

\textup{(1)}~$\KS(b|a,f)=O(\log N)$;

\textup{(2)}~$\KS(b|c,f)=\min\{\KS(a|c),\KS(b|c)\}+O(\log N)$.
\end{thm}

Условие~(1) говорит, что сообщение~$f$ для посвящјнных (знающих
слово~$a$) содержит всю необходимую информацию о~$b$, а
условие~(2) говорит, что для непосвящјнных (знающих только~$c$)
оно содержит минимально возможную информацию о~$b$.

\textbf{Замечание}. С первого взгляда может показаться, что для
доказательства теоремы~\ref{basic} достаточно в качестве $f$
взять кратчайшее описание $b$ при известном $a$ --- дескать,
тогда в нјм не будет ``лишней информации''. Но это не так:
если $a$ и $b$~--- независимые случайные слова длины~$n$, то~$b$
является таким кратчайшим описанием, но не годится в
качестве~$f$ (поскольку $\KS(b)$ и $\KS(a)$ близки к~$n$, а
$\KS(b|b)$ близко к нулю). В этом примере можно взять в
качестве~$f$ побитовую сумму $a\oplus b$ слов~$a$ и~$b$ по
модулю $2$ --- она позволяет знающему $a$ восстановить~$b$, но
$\KS(b|f)\approx n$.

Воспользовавшись теоремой о простом условном
описании~\cite{muchnik-codes}, можно аналогичным способом
доказать утверждение теоремы~\ref{basic}. Согласно теореме о
простом условном описании, существует слово $b'$, которое
является минимальным описанием $b$ при известном $a$, простым
относительно $b$. Это означает, что $\KS(b'|b)=O(\log N)$,
$\KS(b|a,b')=O(\log N)$ и длина $b'$ равна $\KS(b|a)+O(\log N)$.
Рассмотрим также слово $a'$, которое является минимальным
описанием $a$ при известном $b$, простым относительно~$a$, и
уравняем его по длине с $b'$ (отрезав лишние биты или добавив
нули, в зависимости от того, у какого из слов $a$ и $b$ больше
сложность), получится $a''$. Теперь можно взять в качестве~$f$
слово $a''\oplus b'$.

Зная $f$ и $a$, мы можем получить $a'$ и затем $a''$ без
дополнительной информации (точнее, с логарифмической
дополнительной информацией), после чего можно получить $b'$ и
затем $b$. Несложно проверить также, что
$\KS(b|f)=\min\{\KS(a),\KS(b)\}$ с логарифмической точностью.

В самом деле, с логарифмической точностью
        \begin{multline*}
\KS(b|f)=\KS(b,f|f)=\KS(b,b',f|f)=\KS(b,a''|f)\ge \\ \ge
\KS(b,a'')-\KS(f)\ge\KS(b,a'')-|f|=\KS(b,a'')-\KS(b|a)
        \end{multline*}
Слова $a'$ и $b$ независимы (имеют логарифмическую взаимную
информацию), поэтому слово $a''$ (которое получается из $a'$ с
логарифмической сложностью) также независимо с $b$. Поэтому
нижнюю оценку можно продолжить как $\KS(b)-\KS(b|a)+\KS(a'')$,
что равно $\min\{\KS(a),\KS(b)\}$. (Конец альтернативного
доказательства теоремы~\ref{basic}).

Преимущество этого рассуждения в том, что слово $f$ получается
полиномиальной (от $N$) длины и даже минимально возможной длины
$\KS(b|a)$. (В другом доказательстве слово $f$ было функцией с
экспоненциально большой областью определения.) Другими словами,
мы доказали такое утверждение:

\begin{thm}\label{th:1bis}
Для любых двух слов $a$ и $b$ длины не более $N$ найдјтся слово
$f$ длины $\KS(b|a)$, для которого
        $$
\KS(b|f,a)=O(\log N)
        $$
и
        $$
\KS(b|f)=\min\{\KS(a),\KS(b)\}+O(\log N).
        $$
\end{thm}

С другой стороны, недостаток этого рассуждения в том, что оно не
обобщается (по крайней мере непосредственно) на случай
дополнительной информации $c$ (то есть не позволяет доказать
теорему~\ref{relativized}). Пусть опять $a$ и $b$~--- случайные
независимые слова одинаковой длины $2n$, а $a_1$, $a_2$ [$b_1$,
$b_2$]~--- первая и вторая половины слова~$a$ [соответственно
$b$]. Пусть слово $c$ состоит из двух половин $a_1\oplus
a_2\oplus b_1$ и $a_2\oplus b_1\oplus b_2$. Тогда
$\KS(a|c)=\KS(a,c|c)=\KS(a,b|c)=2n$, $\KS(b|c)=2n$, но
$\KS(b|c,a\oplus b)=0$.

В следующем разделе мы дадим комбинаторное доказательство
существования короткого сообщения $f$, которое содержит всю
информацию о $b$ для знающих $a$, но не для знающих только $c$.

\section[\texorpdfstring{Комбинаторное построение
описания минимальной сложности}
{A combinatorial construction of a low complexity description}]
{Комбинаторное построение описания\\ минимальной сложности}

Сейчас мы докажем, что если слово $a$ содержит достаточно много
информации (точнее говоря, если
$\KS(a|c)\ge\KS(b|c)+\KS(b|a)+O(\log N)$), то существует
сообщение $f$, которое обладает указанными в
теореме~\ref{relativized} свойствами и имеет сложность
$\KS(b|a)+O(\log N)$. Для этого мы используем следующую
комбинаторную лемму.
(Через $\mathbb{B}^k$ мы обозначаем множество всех $k$-битовых слов.)

\subsection*{Комбинаторная лемма}

\textbf{Лемма}. \textit{Для любых натуральных $m$ и $n\ge m$
существует семейство $\mathcal{F}$ функций вида
$\mathbb{B}^n\to\mathbb{B}^m$, состоящее из $2^m\poly(n)$
функций и
обладающее таким свойством: для любого слова $b\in\mathbb{B}^m$
и для любого подмножества $\mathcal{F}'$, содержащего не менее
половины всех функций из $\mathcal{F}$, множество точек со
второй координатой $b$, не покрытых графиками функций из
$\mathcal{F}'$, содержит не более $O(2^m)$
 точек.}

\begin{figure}[h]
\begin{center}
   \includegraphics[scale=1]{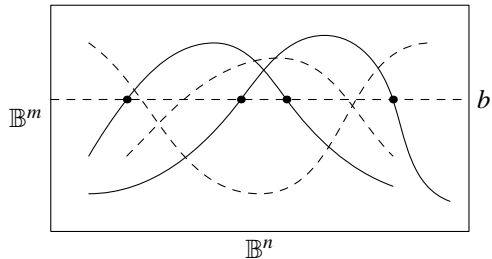}
\end{center}
\caption{Часть функций (до~$50\%$) из $\mathcal{F}$ выброшены;
тем не менее графики оставшихся покрывают любую горизонталь,
не считая $O(2^m)$ точек.}\label{fig}
\end{figure}

Формально свойство семейства $\mathcal{F}$, о котором идјт речь
в лемме (рис.~\ref{fig}), можно записать так:
        $$
\forall b\,\forall \mathcal{F}'\subset\mathcal{F}
\bigl[ |\mathcal{F}'|\ge \frac{1}{2}|\mathcal{F}| \ \Rightarrow \
|\{a\in\mathbb{B}^n\mid f(a)\ne b \text{ для всех $f\in\mathcal{F}'$}\}|=O(2^m)
\bigr]
        $$
(Заметим также, что условие $n\ge m$ излишне: при $n<m$
утверждение леммы выполнено очевидным образом, поскольку
количество всех $a$ есть $O(2^m)$.)

Идея применения леммы: функции, не несущие информации о $b$,
составляют в любом простом семействе большинство, и потому среди
них найдјтся функция, переводящая $a$ в $b$, если только $a$ не
принадлежит малому множеству исключений (что будет противоречить
предположению об~$a$, так как эти исключения имеют малую
сложность).

\begin{proof}
Используя вероятностный метод, возьмјм случайное семейство из
$2^t$ случайных функций $\varphi_1,\ldots,\varphi_{2^t}$. (Все
эти функции независимы и равномерно распределены в множестве
всех функций $\mathbb{B}^n\to\mathbb{B}^m$; точное значение
параметра~$t$ мы выберем позже.) Оценим вероятность того, что
этот набор не обладает требуемым свойством. Это означает, что
найдутся
        \begin{itemize}
\item $b\in\mathbb{B}^m$;
\item множество $S\subset\mathbb{B}^n$ из
$s\cdot 2^m$ элементов (константу~$s$ мы выберем позже);
\item  множество $I\subset\{1,2,\ldots,2^t\}$, содержащее половину всех
индексов,
        \end{itemize}
для которых
        $$
\varphi_i(a)\ne b \text{ для любого $a\in S$ и для любого $i\in I$} \eqno(*)
        $$
Оценим вероятность этого события. Количество различных $b$ есть
$2^m$, количество различных $I$ не превосходит $2^{2^t}$, а
количество различных $S$ не превосходит $(2^n)^{s2^m}$. При
фиксированных $b$, $S$ и $I$ вероятность выполнения условия
$(*)$ есть
        $$
\left( 1 - \frac{1}{2^m}\right)^{2^{t-1}\cdot s2^m}
        $$
(каждая из $2^{t-1}$ функций с номером в $I$ в каждой точке
$a\in S$ не попадает в~$b$ с вероятностью~$(1-1/2^m)$). Всего
для вероятности получаем верхнюю оценку
        $$
2^m\cdot 2^{2^t} \cdot 2^{ns2^m} \cdot
\left( 1 - \frac{1}{2^m}\right)^{2^{t-1}\cdot s2^m},
        $$
и надо показать, что при правильно выбранных значениях
параметров это выражение меньше единицы.
Учитывая, что $(1-1/2^m)^{2^m}\approx 1/e$ (с гораздо большей
точностью, чем нужно нам), можно переписать это как
        $$
2^{m+2^t} \cdot 2^{ns2^m} \cdot
(1/e)^{s2^{t-1}}.
        $$
Основную роль играют члены, которое содержат $2^t$ и $2^m$ в
показателе степени (поскольку $2^t,2^m\gg m,n,s$). Нам нужно,
чтобы малый последний сомножитель перевесил два остальных.
Разобьјм последний сомножитель на две части $(1/e)^{s2^t/4}$,
и сравним эти части с первым и вторым сомножителем. Нам нужно, чтобы
        $$
2^{m+2^t} \cdot
(1/e)^{s2^{t}/4} < 1
        $$
и
        $$
2^{ns2^m} \cdot
(1/e)^{s2^{t}/4}<1.
        $$
Первое неравенство будет выполнено, если взять константу~$s$
достаточно большой. Выполнения второго неравенства
(в котором степени можно сократить
на~$s$) легко добиться, положив $2^t=2^m\poly(n)$.
\end{proof}

\subsection*{Основной результат}

Теперь мы можем дать точную формулировку и доказательство основного
результата:

\begin{thm}\label{thm:main}
Найдјтся такая константа $C$, что для любых слов $a,b,c$ длины
не более $N$, для которых
        $$
\KS(a|c) \ge \KS(b|c) + \KS(b|a) + C\log N,
        $$
существует слово $f$ длины не более $\KS(b|a)+C\log N$,
для которого
$\KS(b|a,f)\le C\log N$ и
$\KS(b|c,f)\ge \KS(b|c)-C\log N$.
\end{thm}

Условие теоремы означает, что имеющееся в распоряжении
получателя слово $a$ содержит достаточно информации,
отсутствующей у ``противника'' (знающего слово $c$);
заключение говорит, что слово $f$ позволяет восстановить $b$ по
$a$ и имеет (среди таких слов) минимально возможную длину, а
также не упрощает восстановление $b$ по $c$. (Во всех выражениях
$O(\log N)$ использована одна и та же константа~$C$, но
поскольку с ростом $C$ утверждение становится слабее, это не
принципиально.)

\begin{proof}
Применим теорему об условном кодировании~\cite{muchnik-codes} и
найдјм слово $b'$ длины $\KS(b|a)$, для которого сложности
$\KS(b|b',a)$ и $\KS(b'|b)$ равны $O(\log N)$. Применим
комбинаторную лемму и построим семейство отображений
$\mathbb{B}^n\to\mathbb{B}^m$ с указанными в ней свойствами,
взяв $m$ равным длине $b'$, то есть $\KS(b|a)$, а $n$ равным
длине $a$. Без ограничения общности можно считать, что семейство
$\mathcal{F}$ является простым, то есть имеющим сложность
$O(\log N)$, поскольку при данных $m$ и $n$ семейство функций с
нужными свойствами можно искать перебором.

Большинство функций в этом семействе (как и в любом другом
простом семействе) не сильно упрощают задание $b$ при известном
$c$ (имеют малое значение разности $\KS(b|c)-\KS(b|c,f)$). В
самом деле, с точностью $O(\log N)$ это можно переписать как
$\KS(f|c)-\KS(f|b,c)$ (по теореме о сложности пары и условной
сложности), среднее значение обоих членов есть логарифм числа
элементов семейства (в данном случае $m+O(\log N)$), поэтому
среднее значение разности есть $O(\log N)$ и остајтся
воспользоваться неравенством Чебышјва.

Оставив в семействе $\mathcal{F}'$ только функции из этого
большинства, мы (по лемме) покроем их графиками пары вида
$\langle a',b'\rangle$ при всех $a'$ длины $n$, кроме $O(2^m)$
``плохих'' $a'$, и нам надо убедиться, что $a$ не попало в
число плохих. Для этого достаточно показать, что
        $$
\KS(a'|c) < \KS(b|c)+\KS(b|a)+O(\log N)
        $$
для всех плохих $a'$. В самом деле, зная $b$ и $c$, а также
$\KS(b|c)$ (последнее число содержит $O(\log N)$ битов, и с
логарифмической точностью им можно пренебречь), можно
перечислять все функции $f$, не попавшие в семейство
$\mathcal{F}'$ (упрощающие задание $b$ при известном $c$), а
потому и точки, не покрытые графиками функций из семейства
$\mathcal{F}'$. (Напомним, что $b'$ можно также восстановить по
$b$ с логарифмической сложностью.) Этих точек по лемме не более
$O(2^m)$, поэтому их сложность при известных $b,c$ не более
$m+O(\log N)$. Получаем
        $$
\KS(a'|b,c)\le \KS(b|a)+O(\log N)
        $$
для всех плохих $a'$, и потому
        $$
\KS(a'|c) \le \KS(a'|b,c)+ \KS(b|c)+O(\log N)\le \KS(b|a)+\KS(b|c)+O(\log N),
        $$
что и требовалось.
\end{proof}

\section{\texorpdfstring{Отрицательный результат и вопросы}
{Negative result and open questions}}

Условие на сложности
в теореме~\ref{thm:main} может показаться искусственным.
Скажем, если слова $a,b,c$ попарно независимы, мы требуем,
чтобы сложность слова $a$ была вдвое больше сложности
слова~$b$, и это выглядит странно (почему агент должен
изначально иметь вдвое больше информации, чем мы хотим передать?
может быть, условия $\KS(a)>\KS(b)$ было бы достаточно?).

В этом разделе мы покажем, что неравенство в условии
теоремы~\ref{thm:main} на самом деле существенно: если этого не
требовать, то может оказаться, что все слова $f$, для которых
выполнено утверждение теоремы~\ref{relativized}, имеют очень
большую длину. Точное утверждение (см. ниже
теорему~\ref{thm:negative}) и его доказательство довольно
длинные, и мы начнјм с более простого рассуждения, которое,
правда, дајт пример с очень большой сложностью слова~$c$. Затем
мы покажем, как можно уменьшить сложность слова~$c$.

Будем строить три слова $a,b,c$, обладающие такими свойствами:
любая не очень длинная (скажем, полиномиальная от сложностей
слов или даже субэкспоненциальная) программа $f$, получающая $b$
из~$a$, значительно упрощает получение $b$ из $c$. В этом
примере слова $a$ и $b$ будут иметь сложность примерно $1{,}3n$
и $n$ соответственно, а их взаимная информация будет близка к
нулю. (Коэффициент $1{,}3$ выбран достаточно произвольно; важно,
что он больше $1$ и меньше~$2$.) Cложность $b$ при известном $c$
будет равна примерно $n$, так что добавление $c$ в качестве
условия не уменьшает сложность $b$. А вот добавление (к
условию~$c$) любой программы $f$, получающей $b$ из $a$,
существенно уменьшает условную сложность $b$: вместо
$\KS(b|c)\approx n$ получается $\KS(b|f,c)\approx 0{,}3n$.

Основная идея конструкции такова: слово $c$ само будет функцией,
переводящей $a$ в $b$ (но при этом без $a$ слово $c$ не содержит
никакой информации о $b$). Пусть нам дана некоторая программа
$f$, которая также переводит $a$ в $b$. Как она помогает задать
$b$ при известном~$c$? Мы знаем, что и $c$, и $f$ переводят $a$
в $b$, поэтому $a$ является одним из решений уравнения
$c(x)=f(x)$. Эти решения можно перечислять, зная $c$ и $f$
(программа $f$ может быть определена не всюду, но это не
мешает). Если решений мало, то можно задать $a$ (а тем самым
и~$b$), указав его порядковый номер в перечислении таких
решений. Этого достаточно при известных $c$ и $f$, и может
требовать меньше битов, чем $\KS(b)$ (которое равно $\KS(b|c)$;
как мы говорили, само по себе $c$ не содержит информации о
$b$).

Покажем, что с большой вероятностью такой пример получится, если
случайно выбрать пару слов $\langle a,b\rangle$ соответствующих длин
($1{,}3n$ и $n$), а в качестве $c$ выбрать случайную функцию (с
аргументами и значениями соответствующих длин), график которой
проходит через $\langle a,b\rangle$. (То же самое распределение вероятностей,
получится, если сначала выбрать случайную функцию, а потом на еј
графике выбрать случайную точку.)

Слова $a$ и $b$ будут иметь нужные сложности и малую взаимную
информацию с большой вероятностью. Чуть сложнее понять, что
сложность $b$ при известном $c$ будет близка к $n$. В самом
деле, типичная функция $c$ с аргументами длины $1{,}3n$ и
значениями длины $n$ принимает большинство своих значений
примерно одинаковое число раз (около $2^{0{,}3n}$),
и потому распределение вероятностей для второй
координаты случайной точки еј графика близко к равномерному, и
большая часть значений будет случайна даже при известной функции
$c$.

Пусть теперь дана некоторая программа $f$, переводящая $a$ в
$b$, причјм она имеет не очень большую длину (существенно
меньшую, чем то, что дајт теорема~\ref{relativized}). Сколько
решений будет иметь уравнение $f(x)=c(x)$? В типичной ситуации
(для данной $f$ и случайной $c$) таких решений будет около
$2^{0,3n}$ (в каждой точке совпадение с вероятностью $2^{-n}$, а
точек $2^{1{,}3n}$); здесь мы предполагаем, что $f$ всюду
определена, но если нет, решений будет только меньше.
Существенно большее число совпадений для данной функции $f$ и
случайно выбранной $c$ очень маловероятно, другими словами, шар
соответствующего радиуса в метрике Хемминга имеет очень малую
вероятность. Поскольку таких шаров не очень много, то и их
объединение будет иметь малую вероятность, и с большой
вероятностью случайная функция $c$ не попадјт ни в один из этих
шаров. Это значит, что для такой функции уравнение $c(x)=f(x)$
для любой не очень длинной программы $f$ будет иметь лишь
немного более $2^{0{,}3n}$ решений, и потому сложности
$\KS(a|f,c)$ и $\KS(b|f,c)$ будут лишь чуть больше $0{,}3n$, как
мы и обещали.

Мы не проводим аккуратно соответствующие оценки, так как хотим
доказать более сильное (и сложное) утверждение. А именно, мы
хотим найти слово $c$ сравнительно небольшой сложности (а не
экспоненциальной сложности, как в изложенной конструкции);
сложность слова $c$ будет примерно на $\KS(b)$ больше, чем
сложность программ $f$, для которых оно будет контрпримером.
(Это не удивительно: чем больше программ $f$ мы должны
``опровергнуть'', тем большая свобода в выборе $c$ требуется.)

Идея конструкции~--- выбор точки на графике случайной функции
$c$~--- сохраняется, только теперь $c$ будет выбираться
не среди всех функций (с аргументами и значениями нужной длины),
а среди функций некоторого семейства $\mathcal{C}$. Мы
сформулируем комбинаторные свойства, которым должно
удовлетворять это семейство, докажем (вероятностно), что такое
семейство существует, заключим отсюда, что существует простое
семейство с этими свойствами (перебор), а затем докажем, что для
большинства точек $a,b$ найдјтся функция $c$ из семейства,
которая удовлетворяет нужным требованиям. (Таким образом, мы
докажем не просто существование тройки $\langle a,b,c\rangle$, а более сильный
факт: для большинства пар $\langle a,b\rangle$ существует $c$.) Ограничивая
размер семейства, мы тем самым ограничиваем сложность слова~$c$.

Нужное нам комбинаторное утверждение мы сформулируем в виде леммы.
Пусть даны множества $A$ и $B$.
Будем говорить, что некоторое семейство $\mathcal{F}$ функций вида $A\to B$
\emph{бракует} функцию $c\colon A\to B$,
если при каком-нибудь $f\in\mathcal{F}$ размер множества
$\{a\colon c(a)=f(a)\}$ больше $4|A|/|B|$
(число корней более чем в $4$ раза превосходит ``ожидаемое'').
Пусть $\mathfrak{H}$~--- набор семейств функций из $A$ в $B$,
параметризованный элементами $b$
(то есть $\mathfrak{H}(b)$ при любом $b\in B$
является семейством функций вида $A\to B$).
Скажем, что функция $c$ \emph{покрывает} пару
$\langle a,b\rangle\in A\times B$ при данных $\mathcal{F}$ и $\mathfrak{H}$,
если $c(a)=b$, при этом $c$ не забраковано семейством~$\mathcal{F}$
и $c\notin \mathfrak{H}(b)$.

\begin{lem}
Пусть $A$ и $B$~--- непустые множества двоичных слов,
причјм $B$ содержит не менее двух элементов и $|A|\ge 16|B|$.
Пусть также даны числа $\varepsilon\ge 4|B|/|A|$ и $\Phi\le 2^{|A|/4|B|}$.
Тогда найдјтся такое семейство $\mathcal{C}$ функций вида $A\to B$ размера
\textup(мощности\textup)
	$$
\max\left\{\frac{20|B|}{\varepsilon},\: \frac{6\Phi\log_2|B|}{\varepsilon},\:
       6\Phi|B|\log_2|B|\right\},
	$$
что для любого семейства $\mathcal{F}$ размера не более $\Phi$
и для любого набора $\mathfrak{H}$ семейств,
каждое из которых имеет размер не более четверти $|\mathcal{C}|$
\textup(то есть $|\mathfrak{H}(b)|\le|\mathcal{C}|/4$ при всех $b\in B$\textup)
не более $\varepsilon$-доли всех пар из $A\times B$ не покрыто ни одним
$c\in \mathcal{C}$ при данных $\mathcal{F}$ и $\mathfrak{H}$.
\end{lem}

Формально утверждение леммы выглядит так
(мы опускаем условия на размеры
$\mathcal{C}$, $\mathcal{F}$ и $\mathfrak{H}(b)$):
        \begin{multline*}
\exists \mathcal{C}
\ \forall \mathcal{F},\mathfrak{H}\\
\biggl|\biggl\{\langle a,b\rangle\colon
\forall c\:
\biggl[ (c(a)=b) \Rightarrow \bigl[(c\in \mathfrak{H}(b))\ \lor\
            (\exists f\in \mathcal{F}\:
                  |\{x:f(x)=c(x)\}|\ge{\textstyle\frac{4|A|}{|B|}}
)\bigr]\biggr]
\biggr\}\biggr|\le\\ \le \varepsilon|A||B|\,.
        \end{multline*}

Смысл этой леммы (как она применяется) такой.
Без ограничения общности можно предполагать, что семейство $\mathcal{C}$ простое (как обычно, можно взять первое в некотором естественном порядке семейство с нужными свойствами).
Возьмјм в качестве $\mathcal{F}$ семейство всех функций,
имеющих простые программы (точнее, их продолжений, если функции частичны),
а в качестве $\mathfrak{H}(b)$~--- семейство всех функций,
простых относительно $b$.
Для пар $\langle a,b\rangle$, не входящих в $\varepsilon$-долю ``плохих'',
найдјтся покрывающая их функция $с$ из семейства $\mathcal{C}$.
Эта функция (точнее, еј индекс в $\mathcal{C}$) и будет нужным контрпримером
(информацией у противника).
Если в дополнение к $c$ противник получает простую программу $f$,
переводящую $a$ в $b$, то задание $a$ и тем самым $b$ упрощается:
достаточно указать порядковый номер $a$ в перечислении всех решений уравнения
$f(x)=c(x)$.
Выбранное нами $\mathfrak{H}$, с другой стороны, гарантирует,
что $c$ сама по себе независима с $b$
(имеет близкую к максимальной сложность даже при известном $b$).
Подробнее мы скажем про это дальше, а пока докажем лемму.

\begin{proof}
В качестве $\mathcal{C}$ возьмјм случайное семейство функций нужного размера.
(Будем считать, что  функции из $\mathcal{C}$ нумеруются числами
от $1$ до $|\mathcal{C}|$, и для каждого номера $i$ и каждой точки $a\in A$
мы независимо определяем значение $i$-й функции на $a$,
выбирая равновероятно один из элементов $B$.)
Докажем, что вероятность того, что $\mathcal{C}$ ``неудачное'',
то есть не обладает требуемым свойством, меньше~$1$.

Оценим вероятность при фиксированном семействе $\mathcal{F}$.
Нам нужно, чтобы для любого набора семейств $\mathfrak{H}$
(с указанными в лемме ограничениями~--- все семейства не более четверти
$\mathcal{C}$) для почти всех пар $\langle a,b\rangle$ нашлась функция из
$\mathcal{C}$, проходящая через точку $\langle a,b\rangle$,
не забракованная семейством $\mathcal{F}$ и не входящая в $\mathfrak{H}(b)$.
Заметим, что в определении забракованной функции не участвует $\mathcal{C}$:
зная $\mathcal{F}$, мы уже знаем, какие функции будут забракованы.
Возможны два случая: забракованных функций много
(скажем, больше четверти всех функций из $\mathcal{C}$)
и забракованных функций мало (меньше четверти).
Во втором случае мы можем присоединить забракованные функции
ко всем семействам $\mathfrak{H}(b)$,
и размер этих семейств останется небольшим
(не больше половины $|\mathcal{C}|$).

Другими словами, для данного семейства $\mathcal{F}$ неудачность $\mathcal{C}$
покрывается объединением следующих двух событий:
\begin{enumerate}
\item $\mathcal{F}$ бракует не меньше четверти  функций из $\mathcal{C}$;
\item для некоторого набора $\mathfrak{H}$,
в котором все семейства $\mathfrak{H}(b)$ имеют размер не более
\emph{половины} $|\mathcal{C}|$,
доля пар из $A\times B$,
через которые не проходит ни одна функция $c\in \mathcal{C}$,
не входящая в $\mathfrak{H}(b)$, превышает $\varepsilon$.
\end{enumerate}

Нам нужно, чтобы сумма вероятностей этих двух событий,
умноженная на количество возможностей для $\mathcal{F}$, была меньше $1$.
Докажем, что вероятность каждого из них меньше $1/2$,
делјнной на $(|B|^{|A|})^{\Phi}$
(последнее выражение~--- верхняя оценка на количество возможных семейств
$\mathcal{F}$ размера не более $\Phi$).

Первое событие запишем в таком виде:
\emph{существует такое подсемейство $\mathcal{C}'\subset \mathcal{C}$
размера $|\mathcal{C}|/4$,
что для всех $c\in \mathcal{C}'$ найдјтся такое подмножество $A'\subset A$
размера $4|A|/|B|$ и такая функция $f\in \mathcal{F}$, что
$f(a)=c(a)$ для всех $a\in A'$.}

Количество возможных $\mathcal{C}'$ оценим сверху общим числом подсемейств
в $\mathcal{C}$, то есть $2^{|\mathcal{C}|}$.
При фиксированном подсемействе (точнее, для данного множества индексов)
функции с этими индексами выбираются независимо,
поэтому можно оценить вероятность нежелательного события для одной из них
и возвести в степень.
Для оценки количества возможных подмножеств $A'$ заметим,
что число подмножеств размера $r$ в множестве размера~$q$,
то есть $C_q^r$, не превосходит $q^r/r!\le {q^r}/{((r/3)^r)}=(3q/r)^r$.
Для $q=|A|$ и $r=4|A|/|B|$ получаем, что
количество возможных~$A'$ не превосходит $(3|B|/4)^{4|A|/|B|}$.

Таким образом, вероятность первого события не больше
$$
   2^{|\mathcal{C}|}\left(
     \Phi
     \left(\frac{3|B|}{4}\right)^{4|A|/|B|}
     \left(\frac{1}{|B|}\right)^{4|A|/|B|}
   \right)^{|\mathcal{C}|/4}=
   \left(2\Phi^{1/4}
     \left(\frac{3}{4}\right)^{|A|/|B|}
   \right)^{|\mathcal{C}|}\,.
$$

После умножения на $|B|^{|A|\Phi}$
(число возможных семейств $\mathcal{F}$) она остајтся меньше $1/2$,
поскольку по условию леммы $|B|\ge 2$, $|A|\ge 16|B|$,
$\Phi\le 2^{|A|/4|B|}$ и $|\mathcal{C}|\ge 6\Phi|B|\log_2|B|$.
Действительно, из последнего условия следует, что
$|\mathcal{C}|\ge 12$ при $\Phi\ge 1$ (случай пустого $\mathcal{F}$ тривиален)
и $|B|\ge 2$,
поэтому $1+|\mathcal{C}|\le 13|\mathcal{C}|/12$ и из $1\le |A|/16|B|$ следует,
что $1+|\mathcal{C}|\le (13/192)(|A||\mathcal{C}|/|B|)$.
Из $\log_2\Phi\le |A|/4|B|$ следует, что
$(|\mathcal{C}|/4)\log_2\Phi \le (1/16)(|A||\mathcal{C}|/|B|)$.
Наконец, из $|\mathcal{C}|\ge 6\Phi|B|\log_2|B|$ следует, что
$|A|\Phi\log_2|B|\le (1/6)(|A||\mathcal{C}|/|B|)$.
Складывая эти неравенства, замечая, что $19/64 < 1/3 < \log_2 (4/3)$,
и потенцируя обе части по основанию $2$,
после группировки сомножителей получаем требуемое неравенство.

Перейдјм ко второму событию (напомним, оно зависит от $\mathcal{F}$,
которое мы предполагаем фиксированным):
\emph{существует такой набор семейств
$\mathfrak{H}=\{\mathfrak{H}(b)\}_{b\in B}$,
в котором каждое семейство $\mathfrak{H}(b)$
имеет размер не более половины $|\mathcal{C}|$,
и такое множество $U\subset A\times B$ размера $\varepsilon|A||B|$,
что для любой пары $\langle a,b\rangle \in U$
и для любой функции $c\in \mathcal{C}$, не принадлежащей $\mathfrak{H}(b)$,
значение $c(a)$ не равно $b$.}

Здесь нам будет удобно считать, что семейства $\mathfrak{H}(b)$
состоят не из функций, а из их индексов (чисел от $1$ до $|\mathcal{C}|$).
(Событие при этом останется тем же.)

Чтобы оценить вероятность интересующего нас (``второго'') события,
зафиксируем $\mathcal{F}$, $\mathfrak{H}$ и $U$.
Тогда соответствующее событие можно описать как пересечение по всем парам $\langle a,b\rangle$
и по всем $i\notin \mathfrak{H}(b)$ событий $c[i](a)\ne b$
(``функция номер $i$ в точке $a$ не равна $b$'').
Оценка была бы простой, если бы эти события были независимы:
тогда надо было бы возвести $(1-1/|B|)$ в степень,
равную количеству всех троек $\langle i,a,b\rangle$,
то есть $\varepsilon|A|\cdot|B|\cdot|\mathcal{C}|/2$.
(Мы умножили число пар $\langle a,b\rangle\in U$
на количество возможных $i$ для данного $b$.)
При разных $a$ (а также при разных $i$) эти события независимы по построению,
но события $c[i](a)\ne b_1$ и $c[i](a)\ne b_2$ зависимы.
Нас выручает то, что зависимость тут ``в нашу пользу''~--- тот факт,
что мы знаем, что $c[i](a)\ne b_1$, только увеличивает вероятность события
$c[i](a)\ne b_2$ (знаменатель в $1/|B|$ уменьшается на единицу,
аналогично и для большего количества условий).
Формально говоря, можно сгруппировать события с одними и теми же $a$ и $i$
и воспользоваться неравенством $(1-k/|B|)\le (1-1/|B|)^k$,
где $k$~--- число событий в группе.

Таким образом, мы получили верхнюю оценку для вероятности неудачи
(при фиксированных $\mathcal{F}$, $\mathfrak{H}$ и $U$): она не превосходит
	$$
\left(1-\frac{1}{|B|}\right)^{\varepsilon|A||B||\mathcal{C}|/2}\le
2^{-\varepsilon|A||\mathcal{C}|/2}\,.
	$$
Остајтся умножить эту вероятность на количество множеств $U$
(их не больше $2^{|A||B|}$),
на количество наборов $\mathfrak{H}$ (их не больше $(2^{|\mathcal{C}|})^{|B|}$)
и на количество семейств $\mathcal{F}$.
В результате получим
	$$
2^{-\varepsilon|A||\mathcal{C}|/2}
2^{|A||B|}
2^{|\mathcal{C}||B|}
|B|^{|A|\Phi}\,.
	$$
Нетрудно проверить, что это меньше $1/2$ при $|B|\ge 2$,
$\varepsilon\ge 4|B|/|A|$, $|\mathcal{C}|\ge 20|B|/\varepsilon$,
и $|\mathcal{C}|\ge (6\Phi\log_2|B|)/\varepsilon$.
Действительно, $1+|A||B|\le 3|A||B|/2$ при непустом $A$ и $|B|\ge 2$,
поэтому из $|\mathcal{C}|\ge 20|B|/\varepsilon$ следует, что
$1+|A||B|\le (3/40)\varepsilon |A||\mathcal{C}|$.
Из $\varepsilon\ge 4|B|/|A|$ следует, что
$|\mathcal{C}||B|\le (1/4)\varepsilon |A||\mathcal{C}|$.
Наконец, из $|\mathcal{C}|\ge (6\Phi\log_2|B|)/\varepsilon$ следует, что
$|A|\Phi\log_2|B|\le (1/6)\varepsilon |A||\mathcal{C}|$.
Складывая эти неравенства, замечая, что $59/120 < 1/2$,
и потенцируя обе части по основанию $2$,
после группировки сомножителей получаем требуемое неравенство.
\end{proof}

Теперь мы можем применить это комбинаторное утверждение
и доказать обещанный отрицательный результат.

Пусть $\alpha>0$~--- некоторая константа.
Пусть $m$, $n$, $l$~--- натуральные числа,
причјм $n\ge 1$, $m\ge n+4$, $m-\alpha\log_2 m\ge n+2$
и $l+1+\log_2 (l+1)\le 2^{m-n-2}$.
Положим $N=\max\{m,l\}$.

\begin{thm}
\label{thm:negative}

Пусть даны слово $a$ длины $m$ и слово $b$ длины $n$,
причјм
	$$
m+n-\KS^{\mathbf{0}'}(a,b)<\alpha\log_2 m.
	$$
Тогда найдјтся слово $c$ сложности $n+l+O(\log N)$, для которого
	\begin{itemize}
\item $\KS(c|b)=\KS(c) + O(\log N)$;
\item $\KS(b|a,c)=O(\log N)$;
\item для любого $f$, при котором $\KS(f)\le l-\KS(b|a,f)$,
      выполнено и $\KS(b|c,f)\le m-n + \KS(b|a,f) +O(\log N)$.
	\end{itemize}
\textup{(}Константа в $O(\cdot)$ зависит от $\alpha$, но
не от $m$, $n$ и $l$.\textup{)}
\end{thm}

Объясним, каким образом эта теорема доказывает существенность условия
в теореме~\ref{thm:main}.
Из равенства $\KS(c|b)=\KS(c) + O(\log N)$ вытекает, что
$b$ и $c$ независимы и $\KS(b|c)=\KS(b)=n$ (с точностью $O(\log N)$).
Поскольку $\KS(b|a,c)=O(\log N)$, то и $\KS(a|c)\ge\KS(b|c)-\KS(b|a,c)=n$
(с точностью $O(\log N)$).
Отметим ещј, что $\KS(b|a)= n$ (с точностью $O(\log m)$).
Таким образом, если $\KS(b|a,f) = O(\log N)$
для некоторого слова $f$ длины не более $l$, то
   $$
\KS(b|c,f) < \min\{\KS(a|c),\KS(b|c)\} + O(\log N)
   $$
при $m - n < n + O(\log N)$, то есть при $\KS(a) < \KS(b|c) + \KS(b|a)$.

\begin{proof}
Пусть $A$~--- множество всех слов длины $m$,
$B$~--- множество всех слов длины $n$.
Возьмјм $\varepsilon=1/m^\alpha$ и $\Phi=2^{l+1}(l+1)$.
Легко убедиться, что из условий на $n,m,l$ следует,
что $A$, $B$, $\varepsilon$ и $\Phi$ удовлетворяют условиям леммы.
Поэтому семейство $\mathcal{C}$ с указанными в лемме свойствами существует.
Его можно эффективно найти перебором, зная $A$, $B$, $\varepsilon$ и $\Phi$,
и поэтому сложность любого элемента $\mathcal{C}$ не превосходит
логарифма числа элементов в $\mathcal{C}$ плюс $O(\log N)$,
то есть $n+l+O(\log N)$.

Возьмјм в качестве $\mathfrak{H}(b)$ множество
$\{c\in \mathcal{C}\colon \KS(c|b) < \log_2|\mathcal{C}|-2\}$.
Ясно, что $|\mathfrak{H}(b)|\le |\mathcal{C}|/4$.

Семейство функций $\mathcal{F}$ строится так.
Оно будет состоять из  $\Phi$ функций, пронумерованных от $1$ до $\Phi$.
Мы перечисляем все тройки $\langle a,b,f\rangle$,
где $a\in A$, $b\in B$ и $f$~--- слово длины $l$,
для которого $\KS(f) + \KS(b|a,f)\le l$.
При этом некоторые номера функций имеют метки, являющиеся словами длины~$l$.
Когда появляется новая тройка $\langle a,b,f\rangle$,
мы пытаемся доопределить в точке $a$ значением $b$ одну из функций,
уже помеченных меткой~$f$.
Если все они уже определены в точке $a$ (и не равны $b$),
мы выбираем номер, который ещј не помечен, помечаем его словом $f$
и полагаем функцию с соответствующим номером равной $b$ в точке $a$.
Каждый раз свободный (ничем не помеченный) номер найдјтся,
поскольку для каждого $f$ нужно не более $2^{l-\KS(f)+1}$ номеров
(если для некоторого $f$ потребовалось больше, значит, для некоторого
$a$ все $2^{l-\KS(f)+1}$ функций определены и принимают разные значения,
то есть мы перечислили уже $2^{l-\KS(f)+1}$ разных $b$,
для которых $\KS(b|a,f)\le l-\KS(f)$~--- противоречие),
а для всех $f$ нужно не более $\sum_{\KS(f)\le l}2^{l-\KS(f)+1}=
\sum_{k=0}^l\sum_{\KS(f)=k}2^{l-k+1}=\Phi$ номеров.
После того, как все тройки с указанным свойством перечислены,
оставшиеся неопределјнными значения функций доопределяются произвольно.

Рассмотрим множество пар $\langle a,b\rangle$,
не покрытых $\mathcal{C}$ при выбранных $\mathcal{F}$ и $\mathfrak{H}$.
Их количество не превосходит $\varepsilon2^{m+n}$.
С другой стороны, $\mathcal{F}$ и $\mathfrak{H}$
могут быть эффективно построены с оракулом $\mathbf{0}'$,
и по ним эффективно строится множество непокрытых пар,
поэтому для любой непокрытой пары
$\KS^{\mathbf{0}'}(a,b)\le m+n-\alpha\log_2 m$.

Таким образом, для любых слов $a$ и $b$, у которых
$m+n-\KS^{\mathbf{0}'}(a,b)<\alpha\log_2 m$,
найдјтся такое $c\in \mathcal{C}$, что $c(a)=b$, $c\notin \mathfrak{H}(b)$
и для любой $f\in \mathcal{F}$ количество таких $x\in A$, что $c(x)=f(x)$,
не больше $2^{m-n+2}$.

Из $c(a)=b$ следует, что $\KS(b|a,c)=O(\log N)$.

Из $c\notin \mathfrak{H}(b)$ следует, что
$\KS(c|b) \ge \log_2|\mathcal{C}|-2$, то есть $\KS(c)=\KS(c|b)+O(\log N)$.

Осталось оценить $\KS(b|c,f)$ для $f$, у которых $\KS(f)\le l-\KS(b|a,f)$.
Зная слово $f$, будем перечислять те функции из $\mathcal{F}$,
которые помечены $f$.
Одна из них, скажем, $\tilde{f}$ содержит пару $\langle a,b\rangle$
(иными словами, $\tilde{f}(a)=b$),
для еј задания нужно не более ${\KS(b|a,f)+O(\log N)}$ дополнительных битов.
Теперь по $\tilde{f}$ и $c$ можно перечислять множество таких $x$,
что $c(x)=\tilde{f}(x)$.
(Строго говоря, при помощи ${\KS(b|a,f)+O(\log N)}$ битов мы можем задать
только номер $\tilde{f}$ в семействе $\mathcal{F}$,
а не саму конечную функцию.
Однако для перечисления решений уравнения $c(x)=\tilde{f}(x)$
нам достаточно того, что мы можем перечислять пары $\langle x,y\rangle$,
для которых $y=\tilde{f}(x)$, повторяя процесс построения $\mathcal{F}$.)
Это множество содержит $a$ и состоит максимум из $2^{m-n+2}$ элементов, и мы можем задать $a$, затратив ещј $m-n+2$ битов. Итого, $\KS(b|c,f)\le \KS(a|c,f) + O(\log N)\le \KS(b|a,f) + m - n + O(\log N)$, что и требовалось.
\end{proof}

\textbf{Открытые вопросы}.

\begin{enumerate}

\item Можно ли в последней теореме улучшить оценку, выбрав $c$ сложности
$n+O(\log N)$ вместо ${n+l+O(\log N)}$? (В својм докладе
Ан.\,А.\,Мучник говорил, что это верно,
но требует более сложного комбинаторного рассуждения,
которое не было рассказано).

\item Последняя теорема показывает, что если $a$ не намного сложнее $b$,
то при некотором $c$ короткие секретные сообщения невозможны.
В то же время альтернативное доказательство теоремы~\ref{basic}
показывает, что при пустом $c$ короткие секретные сообщения всегда возможны.
Можно ли указать какие-то классы слов $c$,
для которых возможны и для которых невозможны короткие сообщения?

\item Что можно сказать о возможных сложностях
$\KS(f|b)$, $\KS(f|a,b)$, $\KS(f|a,b,c)$ для секретных $f$?

\end{enumerate}